\documentclass[12pt]{elsarticle}
\usepackage{slashed}
\usepackage{epsfig,graphics}
\usepackage{hyperref}
\usepackage{bbm}
\usepackage{mathtools}
\usepackage{amssymb}
\usepackage{mathrsfs}
\usepackage{amsmath}
\usepackage{natbib}

\newlength{\llslash}

\newcommand{\tr}{\rm tr \,}

\newcommand{\CD}{\hat \partial}

\begin{document}

\begin{frontmatter}
\title{On chiral excitations with exotic quantum numbers}
\author[GSI]{Xiao-Yu Guo}
\author[SUT]{Yonggoo Heo}
\author[GSI,TUD]{M.F.M. Lutz}
\address[GSI]{GSI Helmholtzzentrum f\"ur Schwerionenforschung GmbH,\\
Planckstra{\ss}e. 1, 64291 Darmstadt, Germany}
\address[SUT]{Suranaree University of Technology, Nakhon Ratchasima, 30000, Thailand}
\address[TUD]{Technische Universit\"at Darmstadt, D-64289 Darmstadt, Germany}
\begin{abstract}
We consider the flavour sextet of charmed meson resonances with $J^P= 1 ^+$ quantum numbers that is predicted by the leading order chiral Lagrangian with up, down and strange quarks. The effect of chiral correction terms as determined previously from QCD lattice data is worked out. Pole masses in the complex energy plane are derived. The most promising signal from such states accessible in experiments like Belle, LHCb and PANDA are foreseen in the s-wave $\pi \,D^*$ phase shift and the $\eta \,D^*$ invariant mass distribution.  For physical quark masses a rapid variation of the phase shift in between the $\eta \,D^*$ and the $\bar K \,D^*_s$ thresholds is predicted.
\end{abstract}
\end{frontmatter}

\section{Introduction}

What are the relevant degrees of freedom of low-energy QCD in terms of which the 
hadronic excitation spectrum can be understood most efficiently and systematically?
This is the fundamental challenge of modern theoretical physics, as QCD is the only fundamental quantum field theory  that in its non-perturbative sector generates a plethora of structure and phenomena in the femto cosmos of the strong interactions. 
The task is to identify the proper degrees of freedom and then construct an effective field theory description of the low-energy hadronic reaction dynamics in terms of the latter. 

A particularly intriguing challenge in QCD is the reaction dynamics of mesons with non-zero charm quantum number \cite{Lutz:2015ejy}. This is so since already the leading order chiral Lagrangian formulated for charmed meson  fields with $J^P= 0^-$ and $J^P = 1^-$ quantum numbers predicts significant short-range forces that may dynamically generate a flavour sextet of resonance states \cite{Kolomeitsev:2003ac,Hofmann:2003je,Lutz:2007sk}. 
Further exotic forces are implied by the leading order chiral Lagrangian for the coupled-channel  interactions of the Goldstone bosons and  baryons with $J^P= \frac{1}{2}^+$ and $J^P = \frac{3}{2}^+$ quantum numbers \cite{Lutz:2003jw,Kolomeitsev:2003kt}. Such predictions reflect the choice of the relevant degrees of freedom in combination with chiral constraints 
stemming from QCD. To this extent detailed studies of these exotic chiral excitations are instrumental to unravel how QCD forms structures out of gluons and quarks. 

The scalar mesons in such a flavour sextet have been studied in some  detail lately \cite{Liu:2012zya,Cleven:2014oka,Albaladejo:2016lbb,Du:2016tgp,Du:2017zvv,Guo:2018kno}. In order to consolidate the leading order predictions of the chiral Lagrangian it is necessary to consider chiral correction terms. Here significant progress is based on using QCD lattice data, as a constraint on the latter. A coherent picture is emerging that confirms the predictions of the leading order computation \cite{Kolomeitsev:2003ac}. In particular the $\pi\, D$ 
scattering phase shift in the isospin-one-half channel is expected to show a clear signal of that exotic flavour sextet just above the $\eta \,D$ threshold. 

The main focus of the current Letter is a comprehensive study of the axial-vector resonances with non-zero charm content as they are dynamically generated by coupled-channel interaction based on the chiral Lagrangian. Here chiral dynamics predicts also the formation of a flavour sextet 
\cite{Kolomeitsev:2003ac}. In turn one would expect an additional isospin-doublet  
that should not be confused with the already established narrow $D_1(2420)$ and broad $D_1(2430)$ states. The latter states are well described to be the heavy-quark spin partners of the $D_2^*(2460)$ and the $D^*_0(2400)$ respectively \cite{Yan:1992gz,Casalbuoni:1996pg,PDG}. The additional exotic states are to be interpreted as the heavy-quark spin partners of the exotic flavour sextet states with $J^P= 0^+$ quantum numbers as discussed most recently in  \cite{Du:2017zvv,Guo:2018kno}. 

What is the impact of chiral correction terms on the exotic $1^+$ systems?  
Such systems received much less attention so far \cite{Guo:2006rp,Gamermann:2007fi,Du:2017zvv}. Here lattice QCD does not yet provide any direct constraints  on the scattering processes of the Goldstone bosons with the charmed mesons ground states with $J^P= 1^-$ quantum numbers. 
We argue, that nevertheless,  progress is possible  by the use of indirect constraints from lattice QCD simulations. There is a sizeable data base on the 
masses of the charmed mesons with $J^P= 1^-$  quantum numbers at various unphysical quark masses. In our recent work \cite{Guo:2018kno} these QCD lattice data were used and estimates for the desired low-energy constants (LEC) were obtained. 

In this work  we derive in particular the s-wave phase shift for the $\pi\, D^*, \eta \, D^*$ and $\bar K \,D_s^*$ channels at various quark masses based on the set of LEC given in  \cite{Guo:2018kno}. In addition we determine the  pole position of all flavour sextet states in the complex plane. Our predictions can be tested on available QCD lattice ensembles of the HSC \cite{Moir:2016srx}.

\section{Coupled-channel interactions from the chiral Lagrangian}

We recall the chiral Lagrangian formulated in the presence of an anti-triplet of charmed  meson field with  $J^P =1^-$ 
quantum numbers \cite{Yan:1992gz,Casalbuoni:1996pg}. In the relativistic version the Lagrangian was 
developed in \cite{Kolomeitsev:2003ac,Hofmann:2003je,Lutz:2007sk} where the $1^-$ field 
is interpolated in terms of an antisymmetric tensor field $D_{\mu \nu}$.
The terms most relevant for our study read 
\begin{eqnarray}
&& \mathcal{L}=
-(\CD_\mu D^{\mu\alpha})(\CD^\nu \bar D_{\nu\alpha})+  \tilde M^2\,D^{\mu\alpha} \,\bar D_{\mu\alpha}/2
\nonumber\\
&& \quad \; \,+\,\big(2\,\tilde{c}_0-\tilde{c}_1\big)\,D^{\mu \nu }\,\bar{D}_{\mu \nu }\,{\tr}\chi _+
 +\tilde{c}_1\,D^{\mu \nu }\,\chi _+\,\bar{D}_{\mu \nu }
\nonumber\\
&& \quad \; \,-\, \big( 4\,\tilde{c}_2+2\,\tilde{c}_3\big)\,D^{\alpha \beta }\bar{D}_{\alpha \beta }\,{\tr}\big(U_{\mu }U^{\mu \dagger }\big)
+2\,\tilde{c}_3\,D^{\alpha \beta }\,U_{\mu }\,U^{\mu \dagger }\,\bar{D}_{\alpha \beta }
\nonumber\\
&& \quad \;\,-\,\big(2\,\tilde{c}_4+\tilde{c}_5\big)\, ({\CD_\mu }D^{\alpha \beta })\,({\CD_\nu }\bar{D}_{\alpha \beta } )\,{\tr} \big[ U^{\mu }, \,U^{\nu \dagger }\big]_+ /\tilde M^2
\nonumber\\
&& \quad \;\, + \,  \tilde{c}_5 \,({\CD_\mu } D^{\alpha \beta })\,\big[U^{\mu },\, U^{\nu \dagger }\big]_+ ({\CD_\nu }\bar{D}_{\alpha \beta })/\tilde M^2 
\nonumber\\
&& \quad \;\,
-\,4\,\tilde{c}_6\,D^{\mu \alpha }\,\big[U_{\mu },\, U^{\nu \dagger }\big]_-\bar{D}_{\nu \alpha } 
\,,
\label{def-kin}
\end{eqnarray}
where
\begin{eqnarray}
&& U_\mu = {\textstyle \frac{1}{2}}\,e^{-i\,\frac{\Phi}{2\,f}} \left(
    \partial_\mu \,e^{i\,\frac{\Phi}{f}} \right) e^{-i\,\frac{\Phi}{2\,f}} \,, \qquad \qquad 
\nonumber\\
&&    \Gamma_\mu ={\textstyle \frac{1}{2}}\,e^{-i\,\frac{\Phi}{2\,f}} \,\partial_\mu  \,e^{+i\,\frac{\Phi}{2\,f}}
+{\textstyle \frac{1}{2}}\, e^{+i\,\frac{\Phi}{2\,f}} \,\partial_\mu \,e^{-i\,\frac{\Phi}{2\,f}}\,,
\nonumber\\
&& \chi_\pm = {\textstyle \frac{1}{2}} \left(
e^{+i\,\frac{\Phi}{2\,f}} \,\chi_0 \,e^{+i\,\frac{\Phi}{2\,f}}
\pm e^{-i\,\frac{\Phi}{2\,f}} \,\chi_0 \,e^{-i\,\frac{\Phi}{2\,f}}
\right) \,, \qquad 
\nonumber\\
&& \CD_\mu \bar D_{\alpha \beta} = \partial_\mu \, \bar D_{\alpha \beta} + \Gamma_\mu\,\bar D_{\alpha \beta} \,, \qquad \quad   \;\; 
\CD_\mu D_{\alpha \beta} = \partial_\mu \,D_{\alpha \beta}  - D_{\alpha \beta}\,\Gamma_\mu \,.
\label{def-chi}
\end{eqnarray}
The covariant derivative $\CD_\mu$ involves the chiral connection $\Gamma_\mu$, the quark masses enter via the symmetry breaking fields $\chi_\pm$ and
the octet of the Goldstone boson fields is encoded into the $3\times3$ matrix $\Phi$. The parameters $f$ and $\tilde M$  give 
chiral limit value of the pion-decay constant and the masses of the $D^*$ mesons respectively. 

We comment on the role of the first order interaction terms
\begin{eqnarray}
{\mathcal L}^{(1)} &=& 2\,g_P\,\{D_{\mu \nu}\,U^\mu\,(\CD^\nu \bar D)
 - (\CD^\nu D )\,U^\mu\,\bar D_{\mu \nu} \}
\nonumber\\
&-& \frac{i}{2}\,\tilde g_P\,\epsilon^{\mu \nu \alpha \beta}\,\{
D_{\mu \nu}\,U_\alpha \,
(\CD^\tau \bar D_{\tau \beta} )
+ (\CD^\tau D_{\tau \beta})\,U_\alpha\,\bar D_{\mu \nu}) \} \,,
\label{def-gP}
\end{eqnarray}
where we consider in addition the pseudo-scalar charm meson field $D$. The low-energy constant $g_P$ 
can be estimated from the decay of the charged $D^*$-mesons \cite{Lutz:2007sk} with
\begin{eqnarray}
|g_P| = 0.57 \pm 0.07 \,,
\label{value-gP}
\end{eqnarray}
The size of the parameter $\tilde g_P$ in (\ref{def-gP}) can not be extracted from empirical data directly. However, the heavy-quark symmetry of QCD \cite{Yan:1992gz,Casalbuoni:1996pg} leads to 
$\tilde g_P = g_P $. While we find in \cite{Guo:2018kno} that such interaction terms are crucial in order 
to quantitatively reproduce the quark-mass dependence of the pseud-scalar and vector charmed meson masses, they are quite irrelevant in the computation of s-wave scattering phase shifts. This is in line with earlier conclusions \cite{Hofmann:2003je,Lutz:2007sk}.

\begin{table}[t]
\setlength{\tabcolsep}{2.5mm}
\renewcommand{\arraystretch}{1.18}
\begin{center}
\begin{tabular}{l|rrrr}
                                            &  Fit 1     &  Fit 2    & Fit 3    & Fit 4      \\ \hline

$ \tilde M\;\;$ \hfill [GeV]                &  2.0636    &  2.1258   &  2.0923  & 2.0728  \\

$ \tilde c_0$                               &  0.2089    &  0.3080   &  0.2737  &  0.2790  \\
$ \tilde c_1$                               &  0.6406    &  0.9473   &  0.8420  &  0.8583  \\
$ \tilde c_2$                               & -0.6031    & -2.2299   & -1.6630  & -1.3452  \\
$ \tilde c_3$                               &  1.2062    &  4.5768   &  3.3260  &  3.0206  \\
$ \tilde c_4$                               &  0.4050    &  2.0418   &  1.2843  &  0.9528  \\
$ \tilde c_5$                               & -0.8099    & -4.2257   & -2.5685  & -2.2205  \\    \hline
$ \tilde c_6$                               &  0         &  0        &  0       &  0       \\ \hline
$ \tilde g_1\;\;$\hfill [GeV$^{-1}$]        &  0         &  0        &  0.4276  &  0.4407  \\
$ \tilde g_2\;\;$\hfill [GeV$^{-1}$]        &  0         &  0        &  1.0318  &  0.8788  \\
$ \tilde g_3\;\;$\hfill [GeV$^{-1}$]        &  0         &  0        &  0.2772  &  0.2003  \\

\end{tabular}
\caption{The low-energy constants from a fit to the pseudo-scalar and vector D meson masses based on QCD lattice ensembles of
the PACS-CS, MILC, ETMC and HSC as explained in \cite{Guo:2018kno}. Each parameter set
reproduces the isospin average of the empirical D and $D^*$ meson masses from the PDG. 
The value $f = 92.4$ MeV was used \cite{Guo:2018kno}.
}
\label{tab:1}
\end{center}
\end{table}

In Tab. \ref{tab:1} we recall four parameter sets as obtained previously in  \cite{Guo:2018kno}. It should be noted that for the $\tilde c_{2-5}$ the results were obtained by relying on the heavy-quark mass limit in which such parameters can be identified with corresponding low-energy constants $c_{2-5}$. With the four fit scenarios of Tab. \ref{tab:1} a global description of the 
world lattice data on charmed meson masses with $J^P=0^-$ or  $J^P=1^-$ quantum numbers together with the few 
constraints available for scattering processes in the $J^P =0^+$ channels was aimed at \cite{Liu:2012zya,Moir:2016srx}. While Fit 1 and Fit 2 provide good results for all $0^-$ and $1^-$ charmed meson masses on various QCD lattice ensembles as well as the 
s-wave scattering lengths of \cite{Liu:2012zya}, they badly fail at reproducing the latest results from HSC \cite{Moir:2016srx}  in particular the $\eta\, D$ phase shift. While in Fit 1 we rely on the leading order 
large-$N_c$ relations 
\begin{eqnarray}
&&\tilde c_2 \simeq -\frac{\tilde c_3}{2} \,,\qquad \qquad 
\tilde c_4 \simeq -\frac{\tilde c_5}{2}\,,
\label{large-Nc}
\end{eqnarray}
for $\tilde c_{2-5}$, in Fit 2 such relations are not imposed. 

Additional subleading operators of chiral order $Q^3$ were considered in  \cite{Guo:2018kno}. While the latter do not affect the masses of the charmed mesons with $J^P= 0^-, 1^-$ they do affect the considered scattering observables. With Fit 3 and Fit 4  all masses and scattering phase shifts as provided on various lattice ensembles are reproduced accurately. The parameter set of choice is Fit 4 where 
we relaxed the leading order large-$N_c$ constraints (\ref{large-Nc})
on all $Q^2$ parameters $\tilde c_{n}$. It is comforting that the deviations from the leading order large $N_c$ relations are typically small. 

The $Q^3$ counter terms were introduced first in \cite{Yao:2015qia,Du:2017ttu} for the charmed meson fields with $J^P =0^-$ fields only. Here we construct the analogous terms for the $J^P =1^-$ fields, where we consider terms only that are relevant for s-wave scattering and are leading in the heavy-quark mass and large-$N_c$ limit
\begin{eqnarray}
&&\mathcal{L}_3 =  - \,2\,\tilde g_1\, \Big( D_{\mu\nu} \,\big[\chi_-,\,U_\rho \big]_-\CD^\rho\bar D^{\mu\nu} -\CD^\rho D_{\mu\nu}\,\big[\chi_-,\,U_\rho \big]_-\bar D^{\mu\nu} \Big)/\tilde M
\nonumber\\
&&\quad \;
    +\,2\,\tilde g_2\, \Big( D_{\mu\nu} \big[\,U^\sigma, \,[\hat \partial_\rho,\, U_\sigma ]_- + [\CD_\sigma,\,U_\rho ]_- \big]_- \CD^\rho \bar D^{\mu\nu} \Big)/\tilde M + {\rm h.c.}
\nonumber\\
&& \quad \;
    +\,2\,\tilde g_3\,\Big( D_{\alpha\beta}\big[\,U_\mu,\,[\CD_\rho,\,U_\nu]_-  \big]_- \big[\CD^\mu,[\CD^\rho,\CD^\nu]_+ \big]_+\bar D^{\alpha\beta} \Big)/\tilde M^3+ {\rm h.c.} \,.
\label{def-gtilde}
\end{eqnarray}
Our detailed study reveals that in the heavy-quark mass limit it holds $ \tilde g_{n} \sim g_n $ for $n=1,2,3$. In turn we can take over the estimates derived for $g_{1-3}$ from  \cite{Guo:2018kno}. They are included in our Tab.  \ref{tab:1} properly matched to the convention used in (\ref{def-gtilde}). Note that we consider here terms only that contribute to s-wave scatterings.

For the parameter $\tilde c_6$ there is no estimate available yet.  Its leading contributions to the s-wave scattering processes are of order $Q^3$, rather than of order $Q^2$. This follows from the particular tensor structure of that interaction term, which also implies that there is no contribution to any of the charmed meson ground state  masses.

\section{Poles in the complex plane and phase-shifts}

The coupled-channel scattering amplitudes, $ T_{ab}(s)$, are derived following the on-shell reduction scheme developed in 
\cite{Lutz:2001yb,Lutz:2003fm,Lutz:2011xc}. For given quantum numbers $J$ and $P$ it can be obtained from the linear system
\begin{eqnarray}
&& T_{ab}(s) = V^{\rm potential}_{ab}(s) + \sum_{c,d} V^{\rm potential}_{ac}(s)\,J^{\rm loop}_{cd}(s)\,T_{db}(s) \,, 
\nonumber\\
&& \qquad {\rm with }\qquad T_{ab}(s = \mu^2) = V^{\rm potential}_{ab}(s= \mu^2)\,,
 \label{def-Tab}
\end{eqnarray}
in terms of an on-shell projected potential term $V^{\rm potential}_{ab}(s)$ taken from the chiral Lagrangian. The loop functions $J^{\rm loop}_{ab}(s)$ are analytic functions in $s$ over the complex plane with a cutline on the real axis from $ s>(m_a+M_a)^2$ only, where $m_a$ and $M_a$ are the masses of the light and heavy mesons of the channel $a$. The explicit form of the loop functions can be taken from \cite{Kolomeitsev:2003ac}. The scheme is applicable for short-range forces only. As the force turns more long range, the corresponding 
left-hand branch points move towards the s-channel unitarity branch points and therefore start to influence the 
physical region more and more. Since the algebraic ansatz (7)
distorts the discontinuity  along the left-hand cut away from its physical form, such an on-shell approximation 
starts to loose control. Any long-range force that has significant effects on the 
scattering process will obscure the application of (\ref{def-Tab}). 
Fortunately, in our application the potentially trouble causing u-channel exchange forces are sufficiently 
weak in the s-wave amplitudes so that their impact can be ignored. The influence of the u-channel contributions is always much smaller than the residual uncertainties discussed in this work. In most cases the effect on the various phase shifts is below 1$^\circ$.

The constraints from crossing symmetry are considered in terms of a properly dialed matching scale $\mu$. 
The latter should be chosen in a domain where the scattering amplitude can be accessed in perturbation theory. In turn for $s$ in the vicinity of the matching point the amplitude will be consistent with all  constraints set by crossing symmetry. Depending on whether we are interested in $s \gg \mu ^2$
or $s \ll \mu^2$ we may or may not use (\ref{def-Tab}). Only in a domain where the physics is dominated by the s-channel unitarity cuts (\ref{def-Tab}) can be justified. In the other case at  $s \ll \mu^2$ the 
unitarization should be set up in a crossed channel instead and (\ref{def-Tab}) is invalid. 
According to [2] natural values for the matching scale are 
$\mu = M_{D^*}$ or $\mu =M_{D^*_s}$, where the choice depends on the strangeness of the considered coupled-channel system only.

In a first step we compute the poles in the complex plane, where we focus on the exotic sextet channels. 
The analytic continuation of the scattering amplitudes on higher Rieman sheets is straight forward and well documented in the literature (see e.g. \cite{JohnTaylor}). For an $n$-dimensional coupled channel system the specifics of the particular sheet are referred to by $n$ signatures $(\pm, \cdots ,\pm)$, where a $+$ at position $i$ signals that the $i$-th channel unitarity branch cut remains on the real axis on the interval $ \{(m_i+M_i)^2, \infty \}$. In contrast a negative signature $-$ refers to a rotation of the $i$-th unitarity branch cut onto the interval $\{- \infty, (m_i+M_i)^2\}$. It is important to remember, that any such rotation of a cut line does not modify the scattering amplitude in the upper complex plane.

\begin{table}[t]
\setlength{\tabcolsep}{2.5mm}
\renewcommand{\arraystretch}{1.5}
\hspace*{+0.25cm}
\begin{tabular}{c|c|c|c|c}
\hline
            &$\Delta\mu$&$(I,S)= (1,1)$      &$(I,S)= (1/2,0)$    &$(I,S)= (0,-1)$     \\ \hline
WT          & $-0.1$    &   $2.599-0.075i$ &   $2.507-0.036i$ &   $2.497$             \\
            & $ 0$      &   $2.618-0.080i$ &   $2.525-0.036i$ &   $2.484$       \\
            & $+0.1$    &   $2.641-0.093i$ &   $2.545-0.035i$ &   $2.453$             \\ \hline
Fit 1       & $-0.1$    &   $2.626-0.085i$ &   $2.596-0.041i$ &   $2.509-0.111i$ \\
            & $ 0$      &   $2.642-0.094i$ &   $2.603-0.040i$ &   $2.511-0.125i$ \\
            & $+0.1$    &   $2.658-0.108i$ &   $2.611-0.035i$ &   $2.512-0.143i$ \\ \hline
Fit 2       & $-0.1$    &   $2.649-0.239i$ &   $2.659-0.126i$ &   $2.477-0.142i$ \\
            & $ 0$      &   $2.657-0.248i$ &   $2.668-0.121i$ &   $2.475-0.154i$ \\
            & $+0.1$    &   $2.662-0.259i$ &   $2.681-0.111i$ &   $2.471-0.168i$ \\\hline
Fit 3       & $-0.1$    &   $2.566-0.291i$ &   $2.604-0.075i$ &   $2.394-0.119i$ \\
            & $ 0$      &   $2.575-0.301i$ &   $2.625-0.059i$ &   $2.391-0.131i$ \\
            & $+0.1$    &   $2.582-0.312i$ &   $2.647-0.035i$ &   $2.387-0.145i$ \\\hline
Fit 4       & $-0.1$    &   $2.554-0.281i$ &   $2.576-0.072i$ &   $2.392-0.079i$ \\
            & $ 0$      &   $2.564-0.290i$ &   $2.606-0.059i$ &   $2.390-0.095i$ \\
            & $+0.1$    &   $2.573-0.302i$ &   $2.629-0.034i$ &   $2.386-0.113i$ \\ \hline
\end{tabular}
\caption{Pole masses of the $1^+$  meson resonances in the flavour sextet channels. The $(1,1), (1/2,0), (0,-1)$  poles are located on the $(-,+), (-,-,+), (-)$ sheets respectively. All quantities are given in units of GeV. The parameter sets of Tab. \ref{tab:1} are used. With 'WT' we refer to the leading order scenario that relies on the parameter $f = 92.4$ MeV only.}
\label{tab:2}
\end{table}

\begin{figure}[t]
\includegraphics[keepaspectratio,width=0.98\textwidth]{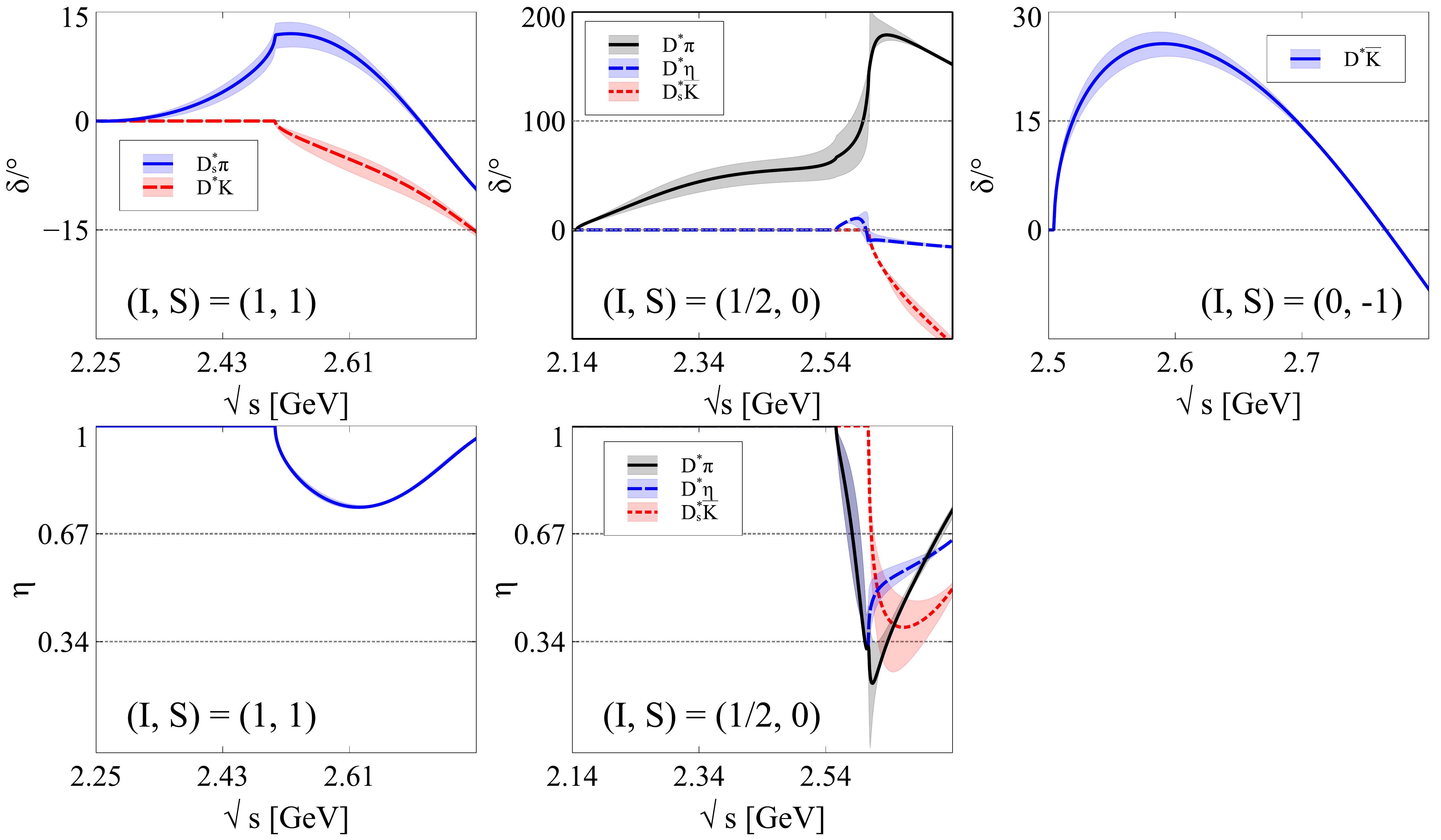}
\includegraphics[keepaspectratio,width=0.98\textwidth]{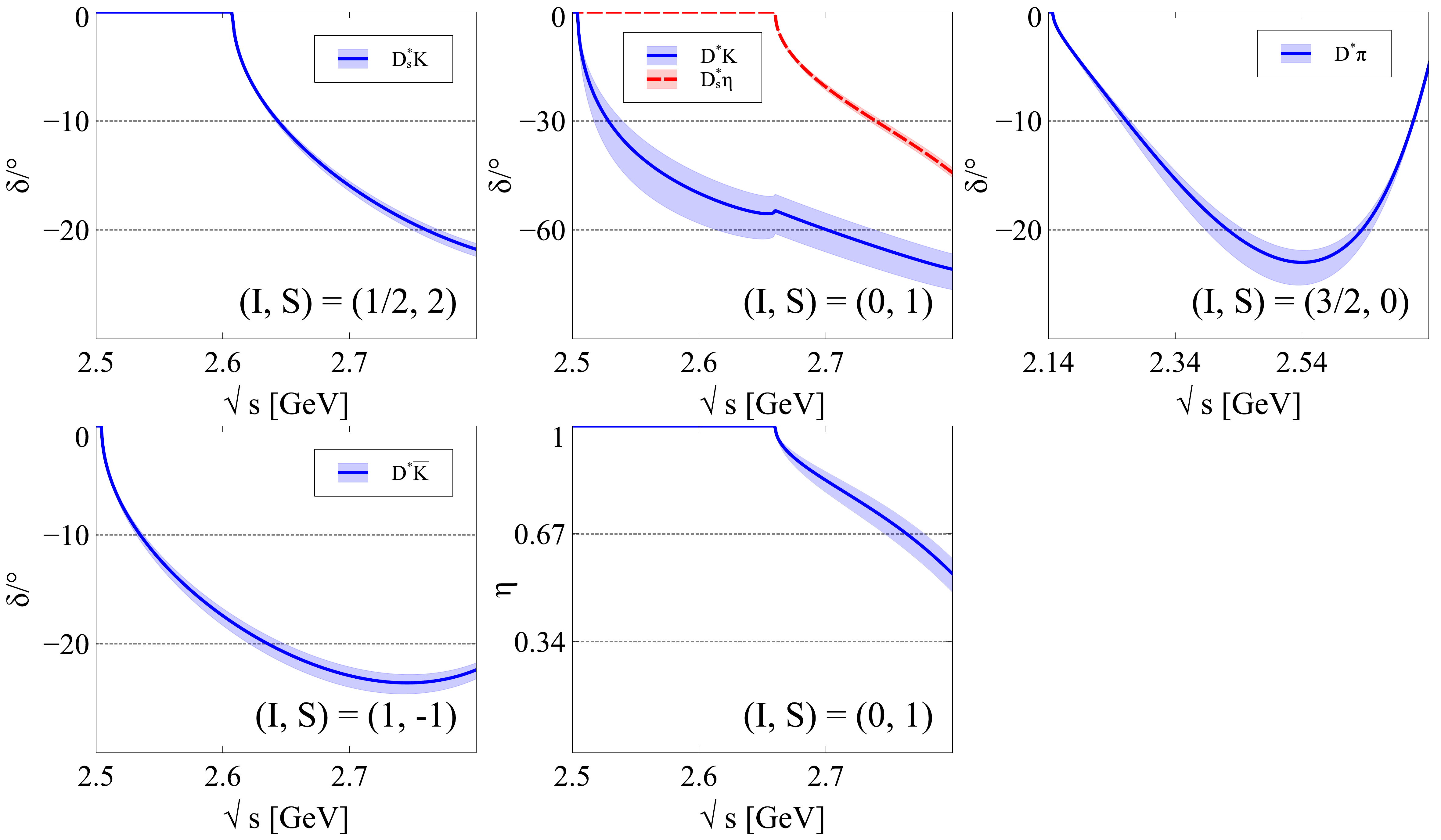}
\vskip-0.05cm
\caption{\label{fig:1} S-wave phase shifts and in-elasticity parameters based on the parameter set Fit 4 in Tab. \ref{tab:1}. The shaded bands are implied by a variation of the matching scales with $|\Delta \mu |<  0.1$ GeV. }
\end{figure}

\begin{figure}[t]
\includegraphics[keepaspectratio,width=0.98\textwidth]{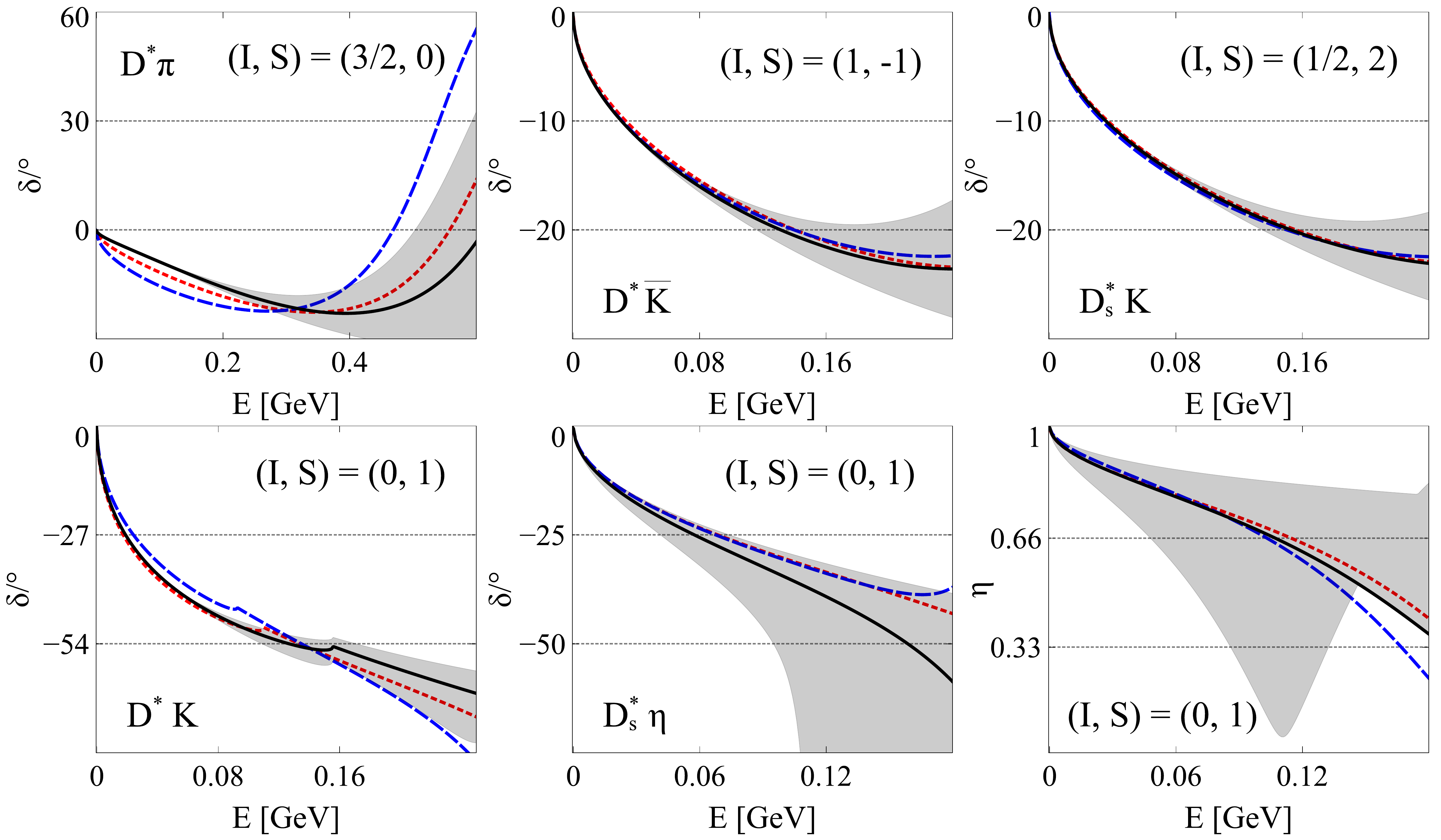}
\includegraphics[keepaspectratio,width=0.98\textwidth]{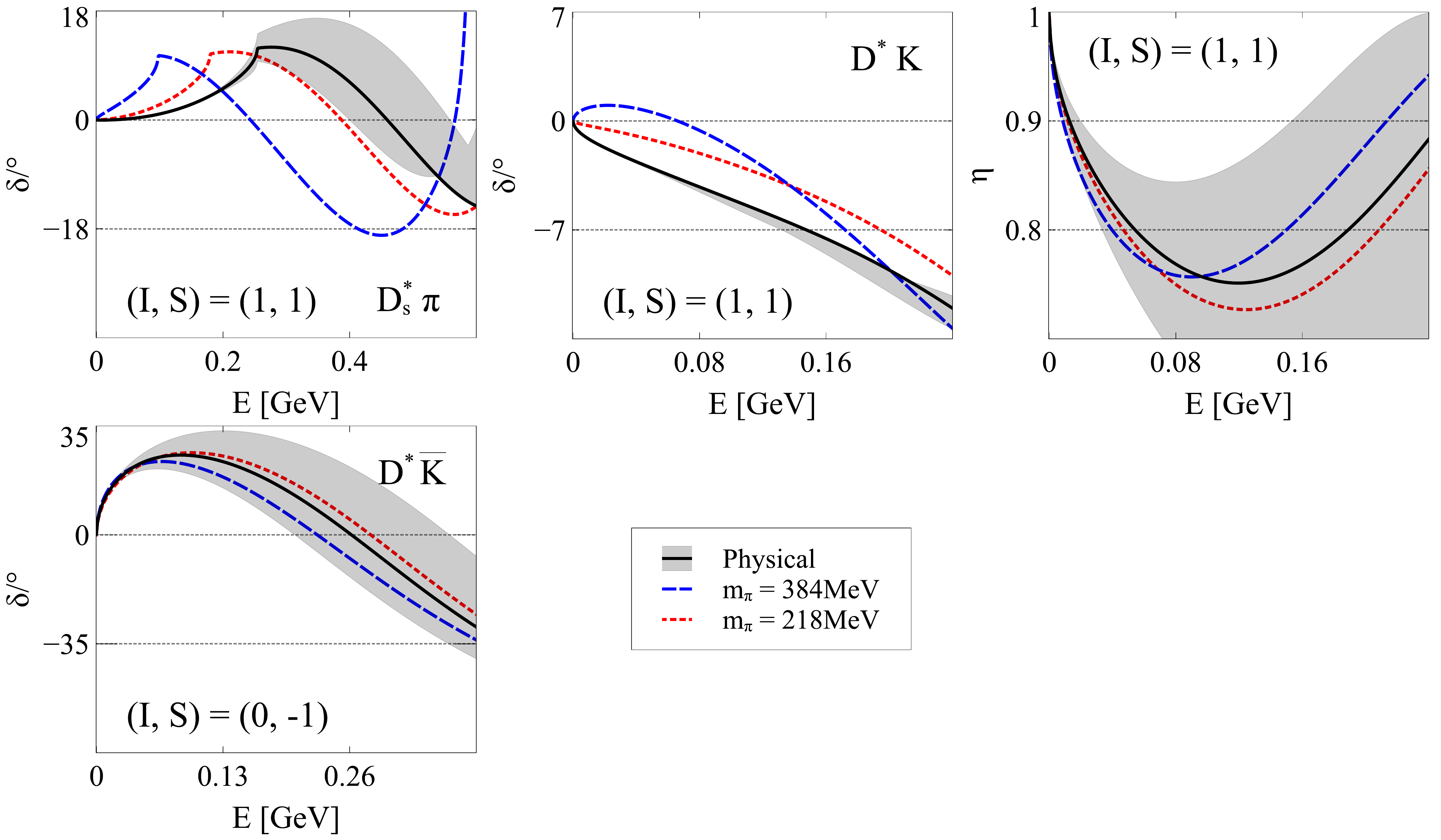}
\vskip-0.05cm
\caption{\label{fig:2} S-wave phase shifts and in-elasticity parameters based on the parameter set Fit 4 in Tab. \ref{tab:1}.  Results are shown for three different pion masses as explained in the text. Error bands are shown only for case where the pion mass takes its physical value. 
The shaded bands are implied by a variation of the unknown LEC  with $|\tilde c_6 |< 1$. The energy $E$ is measured relative to the threshold of the lightest channel always.}
\end{figure}

\begin{figure}[t]
\includegraphics[keepaspectratio,width=0.97\textwidth]{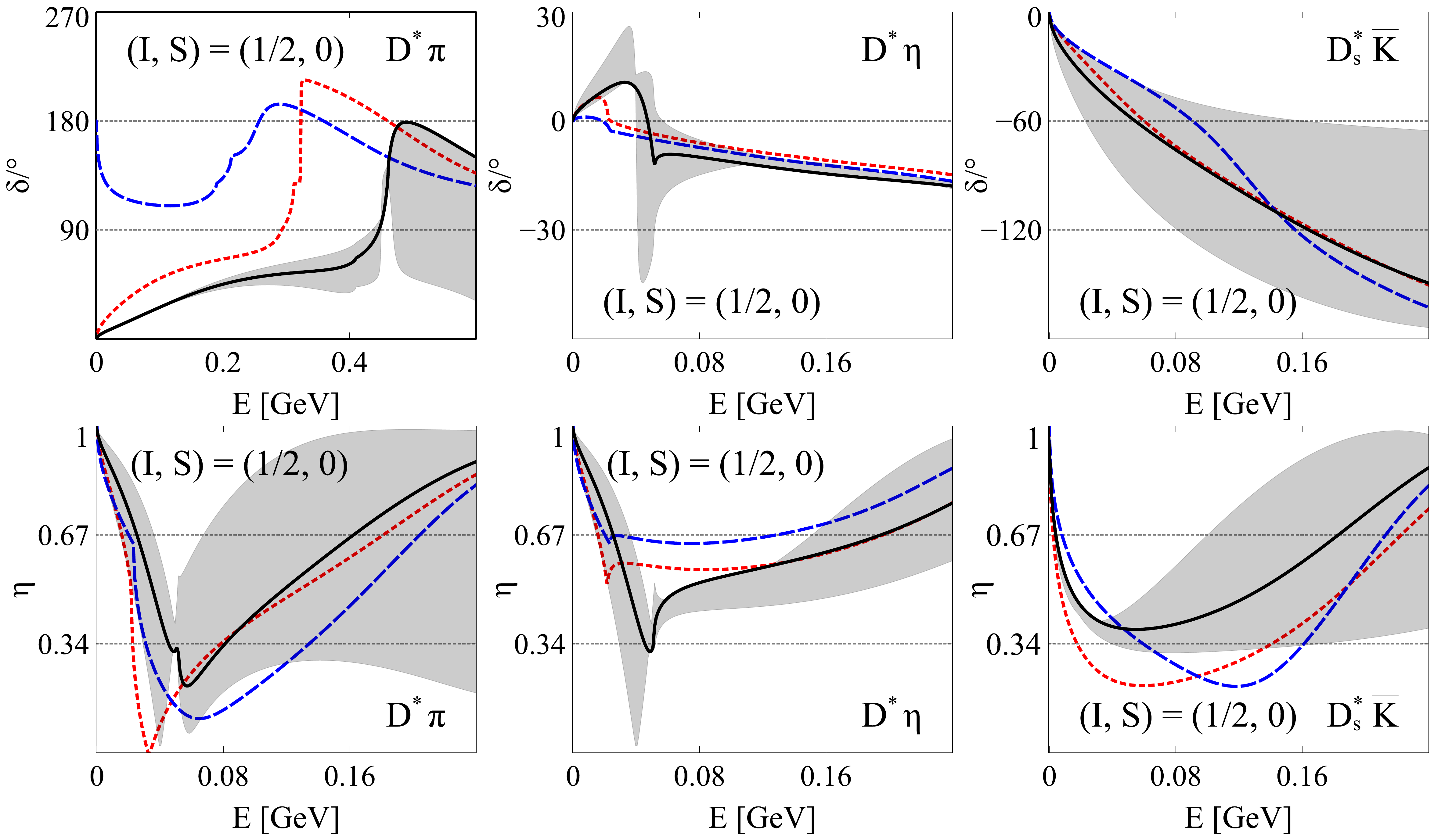}
\vskip-0.05cm
\caption{\label{fig:3} S-wave phase shifts and in-elasticity parameters based on the parameter set Fit 4 in Tab. \ref{tab:1}. Conventions are as in  Fig. \ref{fig:2}. }
\end{figure}

In Tab. \ref{tab:2} we present the pole masses in all flavour sextet channels. We do so for the four parameters sets of Tab. \ref{tab:1}. While there are quantitative differences in the masses  we find a clear signal for the three exotic states for all four parameter sets. An estimate of the theoretical error is attempted in terms of a variation of the matching scales, $\mu$, around  their natural values as introduced in \cite{Kolomeitsev:2003ac}. We use $\Delta \mu =\pm\, 0.1$ GeV throughout this work.
In all cases the most prominent signal of the exotic flavour sextet is seen as a second pole in the $(I,S)=(1/2,0)$ sector. In fact the first pole with a rather large width is associated with a member of the flavour anti-triplet, to which belongs also the well established $D_{s1}(2460)$ state. For our Fit 4 we obtain the masses $M_{D_{1}(2430)} = (2.228_{+1}^{-8}-0.182_{-28}^{+44}\,i)$ GeV and
$M_{D_{s1}(2460)} =2.431_{-36}^{+36} $ GeV where the asymmetric error is implied by  $\Delta \mu =\pm\, 0.1$ GeV. The empirical  mass of the $D_{s1}(2460)$ is reproduced within the estimated natural range of the  matching scale $\mu$.
Note that we do find additional poles not shown in Tab. \ref{tab:2} that are significantly further away from the physical region and therefore have only a very minor influence on the scattering processes. Such poles are outside the energy domain where we would trust our approach.

In case of the first pole with $(I,S)=(1/2,0)$ we observe a strong coupling into the $\pi \,D^*$ and therefore a width ranging from 300-500 MeV. Experimental hints for the existence of such a broad state are documented in \cite{Abe:2003zm,Aaij:2016fma}. Here we feel that it is more significant to directly show the s-wave $\pi \, D^*$ phase shift. To ponder on the pole mass for such a state at this early stage of the theory development we  find misleading since the systematic uncertainties typically get amplified the deeper we go into the complex plane. In Fig. \ref{fig:1} all s-wave phase shifts and in-elasticity parameters are shown for our favourite Fit 4 parameters. The uncertainty bands are implied by a variation of the matching scale $\mu$ by $|\Delta \mu| < 0.1$ GeV. In all cases the uncertainty bands are comfortably small, illustrating the significance of our predictions. Most impressive we find the strong rise in the $\pi D^*$ phase shift above the $\eta \,D^* $. An experimental confirmation would be a direct signal for the existence of such a flavour sextet in QCD \cite{Du:2017zvv}.
We note that analogous predictions hold for the s-wave $\pi \,D$ and $\eta \,D$ phase shifts\cite{Kolomeitsev:2003ac,Albaladejo:2016lbb,Guo:2018kno}. Here our resonance mass  comes at $(2.439_{-32}^{+41} - 0.092_{+3}^{-7}\, i )$ GeV in Fit 4, which is in qualitative agreement only with the value $(2.451_{-26}^{+36} - 0.134_{-8}^{+7}\,i)$ GeV of \cite{Albaladejo:2016lbb}.

In order to offer additional material that permits lattice QCD simulations to scrutinize our predictions 
we provide the additional Fig. \ref{fig:2} and Fig. \ref{fig:3} in which the various phase shifts are detailed at different choices for the pion mass. Here we compare the results at physical quark masses with those that would follow on different lattice ensembles currently used by HSC \cite{Cheung:2016bym}. 
In most cases variations caused by the use of different quark masses are of minor importance. The most striking exception are the s-wave $\pi \,D^*$ phase shifts shown in Fig. \ref{fig:3}. This resembles the strong impact of the choice of quark masses already documented for the corresponding s-wave $\pi \,D$ phase shift in our previous work \cite{Guo:2018kno}.  

It should be noted that at this stage there is yet a further source of uncertainty as a tribute to the so-far unknown LEC $\tilde c_6$. We estimate the influence of a non-vanishing $|\tilde c_6| < 1$ by error bands in Fig. \ref{fig:2} and Fig. \ref{fig:3} associated with the solid lines. Fortunately our main conclusions are not affected.

\section{Summary and outlook}

Based on a set of low-energy parameters determined previously from lattice QCD data we make 
predictions for the s-wave phase shifts of the Goldstone bosons off the flavour anti-triplet of charmed mesons  with $J^P = 1^-$ quantum numbers. Such phase shifts are required in high-precision Dalitz plot analyses of B meson decays, which are on the way for Belle II and LHCb data. 
It is found that the chiral correction terms further consolidate the leading order prediction of the chiral Lagrangian that there exists an exotic flavour sextet in QCD with $J^P = 1^+$ quantum numbers. The most promising signal would be a rapid variation of the s-wave $\pi \,D^*$ phase shift in-between the $\eta \,D^*$ and $\bar K\,D_s^*$ thresholds \cite{Du:2017zvv}. Supplementary results on the pole-positions of the flavour sexet states in the complex energy plane as well as the quark-mass dependence of all s-wave phases shifts are provided. 

It remains to investigate how the already established narrow $D_1(2420)$ state is possibly modifying the detailed structure of our axial-vector state. The fact that the $D_1(2420)$ is outside the effective Lagrangian  approach considered we take as a strong hint that further degrees of freedom have to be taken into account. Following \cite{Lutz:2007sk} we would speculate that the nonet of light vector mesons may play a decisive role. Here the impact of the s-wave $\rho \,D$ channel on the decay width of the  $D_1(2420)$  is expected to be of particular importance. 

\vskip0.3cm
{\bfseries{Acknowledgments}}
\vskip0.3cm
Y. H. acknowledges partial support from Suranaree University of Technology, the Office of the Higher Education Commission under NRU project 
of Thailand (SUT-COE: High Energy Physics and Astrophysics) and  SUT-CHE-NRU (Grant No. FtR.11/2561).


\bibliographystyle{elsarticle-num}
\bibliography{thesis}

\end{document}